\documentclass[10pt,superscriptaddress,prc,aps,twocolumn,showpacs]{revtex4}
\usepackage{amsmath,amssymb}
\usepackage{graphicx}

\newcommand{\Glasgow}[1]
{\affiliation{SUPA School of Physics and Astronomy, University of Glasgow, Glasgow G12 8QQ, United Kingdom}}

\newcommand{\Mainz}[1]
{\affiliation{Institut f\"ur Kernphysik, Johannes Gutenberg-Universit\"at Mainz, D-55099
Mainz, Germany}}

\newcommand{\Bonn}[1]
{\affiliation{Helmholtz-Institut f\"ur Strahlen- und Kernphysik, Universit\"at Bonn, D-53115 Bonn, Germany}}

\newcommand{\Dubna}[1]
{\affiliation{Joint Institute for Nuclear Research, 141980 Dubna, Russia}}

\newcommand{\Pavia}[1]
{\affiliation{INFN Sezione di Pavia, I-27100 Pavia, Italy}}

\newcommand{\GWU}[1]
{\affiliation{The George Washington University, Washington, DC 20052-0001, USA}}

\newcommand{\LPI}[1]
{\affiliation{Lebedev Physical Institute, 119991 Moscow, Russia}}

\newcommand{\Halifax}[1]
{\affiliation{Department of Astronomy and Physics, Saint Mary’s University, Halifax, Nova Scotia B3H 3C3, Canada}}

\newcommand{\Basel}[1]
{\affiliation{Departement f\"ur Physik, Universit\"at Basel, CH-4056 Basel, Switzerland}}

\newcommand{\Tomsk}[1]
{\affiliation{Laboratory of Mathematical Physics, Tomsk Polytechnic University, 634034 Tomsk, Russia}}

\newcommand{\Edinburgh}[1]
{\affiliation{SUPA School of Physics, University of Edinburgh, Edinburgh EH9 3JZ, United Kingdom}}

\newcommand{\INR}[1]
{\affiliation{Institute for Nuclear Research, 125047 Moscow, Russia}}

\newcommand{\Sackville}[1]
{\affiliation{Mount Allison University, Sackville, New Brunswick E4L 1E6, Canada}}

\newcommand{\Regina}[1]
{\affiliation{University of Regina, Regina, Saskatchewan S4S 0A2, Canada}}

\newcommand{\Zagreb}[1]
{\affiliation{Rudjer Boskovic Institute, HR-10000 Zagreb, Croatia}}

\newcommand{\Kent}[1]
{\affiliation{Kent State University, Kent, Ohio 44242-0001, USA}}

\newcommand{\Amherst}[1]
{\affiliation{University of Massachusetts, Amherst, Massachusetts 01003, USA}}

\newcommand{\Bochum}[1]
{\affiliation{Institut f\"ur Experimentalphysik, Ruhr-Universit\"at, D-44780 Bochum,
Germany}}

\newcommand{\UCLA}[1]
{\affiliation{University of California Los Angeles, Los Angeles, California 90095-1547, USA}}

\begin{document}

\title{\boldmath
First measurement of target and beam-target asymmetries in the $\gamma p\to \pi^0\eta p$ reaction}

\author{J.~R.~M.~Annand}\Glasgow  \\
\author{H.~J.~Arends}\Mainz \\
\author{R.~Beck}\Bonn \\
\author{N.~Borisov}\Dubna \\
\author{A.~Braghieri}\Pavia \\
\author{W.~J.~Briscoe}\GWU \\
\author{S.~Cherepnya}\LPI \\
\author{C.~Collicott}\Halifax \\
\author{S.~Costanza}\Pavia \\
\author{E.~J.~Downie}\Mainz \\ \GWU \\
\author{M.~Dieterle}\Basel \\
\author{A.~Fix}\thanks{fix@tpu.ru}\Tomsk \\
\author{L.~V.~Fil'kov}\LPI \\
\author{S.~Garni}\Basel \\
\author{D.~I.~Glazier}\Edinburgh \\ \Glasgow \\
\author{W.~Gradl}\Mainz \\
\author{G.~Gurevich}\INR \\
\author{P.~Hall Barrientos}\Edinburgh \\
\author{D.~Hamilton}\Glasgow \\
\author{D.~Hornidge}\Sackville \\
\author{D.~Howdle}\Glasgow \\
\author{G.~M.~Huber}\Regina \\
\author{V.~L.~Kashevarov}\thanks{kashev@kph.uni-mainz.de}\Mainz \\ \LPI \\
\author{I.~Keshelashvili}\Basel\\
\author{R.~Kondratiev}\INR \\
\author{M.~Korolija}\Zagreb \\
\author{B.~Krusche}\Basel \\
\author{A.~Lazarev}\Dubna \\
\author{V.~Lisin}\LPI \\
\author{K.~Livingston}\Glasgow \\
\author{I.~J.~D.~MacGregor}\Glasgow \\
\author{J.~Mancel}\Glasgow \\
\author{D.~M.~Manley}\Kent \\
\author{P.~P.~Martel}\Amherst \\ \Mainz \\
\author{E.~F.~McNicoll}\Glasgow \\
\author{W.~Meyer}\Bochum \\
\author{D.~G.~Middleton}\Sackville \\ \Mainz \\
\author{R.~Miskimen}\Amherst \\
\author{A.~Mushkarenkov}\Pavia \\ \Amherst \\
\author{A.~Neganov}\Dubna \\
\author{A.~Nikolaev}\Bonn \\
\author{M.~Oberle}\Basel \\
\author{H.~Ortega}\Mainz \\
\author{M.~Ostrick}\Mainz \\
\author{P.~Ott}\Mainz \\
\author{P.~B.~Otte}\Mainz \\
\author{B.~Oussena}\Mainz \\ \GWU \\
\author{P.~Pedroni}\Pavia \\
\author{A.~Polonski}\INR \\
\author{V.~V.~Polyanski}\LPI \\
\author{S.~Prakhov}\UCLA \\
\author{G.~Reicherz}\Bochum \\
\author{T.~Rostomyan}\Basel \\
\author{A.~Sarty}\Halifax \\
\author{S.~Schumann}\Mainz \\
\author{O.~Steffen}\Mainz \\
\author{I.~I.~Strakovsky}\GWU \\
\author{Th.~Strub}\Basel \\
\author{I.~Supek}\Zagreb \\
\author{L.~Tiator}\Mainz \\
\author{A.~Thomas}\Mainz \\
\author{M.~Unverzagt}\Mainz \\
\author{Yu.~A.~Usov}\Dubna \\
\author{D.~P.~Watts}\Edinburgh \\
\author{D.~Werthm\"uller}\Basel \\
\author{L.~Witthauer}\Basel \\
\author{M.~Wolfes}\Mainz \\

\collaboration{A2 Collaboration at MAMI}

\date{today}

\begin{abstract}
The first data on target and beam-target asymmetries for the $\gamma p\to\pi^0\eta p$ reaction
at photon energies from 1050 up to 1450 MeV are presented.
The measurements were performed using the Crystal Ball and TAPS detector
setup at the Glasgow tagged photon facility of the Mainz Microtron MAMI.
The general assumption that the reaction is dominated by
the $\Delta 3/2^-$ amplitude is confirmed.
The data are in particular sensitive to small contributions from other partial waves.
\end{abstract}

\pacs{25.20.Lj, % Photoproduction reactions
      13.60.Le, % Meson production
      14.20.Gk  % Baryon resonances with S=0
      } %

\maketitle

%%%%%%%%%%%%%%%%%%%%%%%%%%%%%%%%%%%%%%%%%%%%%%%%%%%%%%%%%%%%%%%%%%%%%%%%%%%%%%%%%%%%%%
\section{Introduction}\label{Intro}

Photoinduced production of $\pi^0\eta$ pairs is a relatively new topic in particle
physics. Nevertheless, since modern $4\pi$ photon detectors in combination with high
intensity photons beams have become available, a large amount of data, primarily angular
and momentum distributions have been measured. The production of the meson pairs is
sensitive to sequential decays of baryon resonances such as $\Delta^*\to\Delta(1232)\eta
\to N\pi^0\eta$ and $\Delta^*\to N^*(1535)\pi^0 \to N\eta\pi^0$. Hence, $\pi^0\eta$ meson
pair production is a complementary tool to study nucleon and $\Delta$ excitation spectra
providing additional information compared to single $\pi$ or $\eta$ photoproduction.
Remarkably, till now no major disagreements between experimental data and
state-of-the-art model predictions for the $\gamma p\to\pi^0\eta p$ reaction have been
observed. The main reason for this is due to the fact that this reaction seems to be
dominated by just a single partial-wave amplitude. Several independent studies
\cite{Doring,HornEPJ,Horn1940,FOT,FKLO,Aphi,Anis12,W_LM,Gutz2014} agree with the
assumption that the $\gamma p\to\pi^0\eta p$ reaction mainly proceeds via excitation of
the $\Delta 3/2^-$ amplitude with a moderate admixture of positive parity resonances and
generally insignificant contributions from nonresonant Born terms. As a consequence, even
though the models differ from each other in detail, they provide similar results for many
observables. In order to disentangle small components in the reaction amplitude, it is
therefore important to study spin observables which are especially sensitive to
interference terms. The situation is similar to single $\pi^0$ photoproduction at
energies up to $E_{\gamma}= 400  ~\mathrm{MeV}$, which is dominated by the magnetic
$\Delta 3/2^+$ multipole amplitudes due to the excitation of the $\Delta(1232)P_{33}$
resonance or single $\eta$ production close to threshold, which is dominated by the
$N^*(1535)S_{11}$ resonance.

The possibility of model independent partial-wave analysis of a so-called ``complete''
set of measurements is often one of the main motivations given for polarization
measurements. Such a ``complete experiment,'' which is a complex and extensive  task for
single meson photoproduction, is even more difficult for reactions in which two mesons
are emitted. However, in some cases it is possible to study the partial wave content
using a restricted number of observables, making some physically reasonable general
assumptions about the production mechanisms. Some polarization observables for $\gamma
p\to \pi^0\eta p$ were already measured and analyzed in earlier papers
\cite{Aphi,Ajaka,Gutz1,Gutz2,DorMeiss}. Here we report the first measurements of
asymmetries obtained using a transversely polarized proton target. Our main objectives
are to check the consistency of the experimental target and beam-target asymmetry data
with the dominant $\pi^0 \eta$ production mechanism and to investigate small
contributions from partial waves other than the dominant $\Delta 3/2^-$ amplitude.

%%%%%%%%%%%%%%%%%%%%%%%%%%%%%%%%%%%%%%%%%%%%%%%%%%%%%%%%%%%%%%%%%%%%%%%%%%%%%%%%%%%%%%
\section{General formalism}

\begin{figure}
\begin{center}
\resizebox{0.4\textwidth}{!}{%
\includegraphics{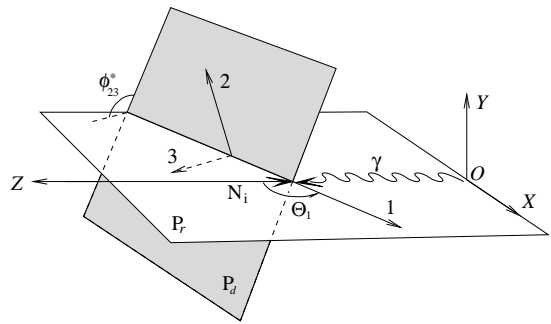}}
\caption{Definition of the coordinate systems used in the present work. The azimuthal
angle $\phi_{23}^*$ is defined in the center-of-mass system (cm) system of particles 2
and 3 with the $z$ axis opposite to the momentum of particle 1 and $y$ axis parallel to
$OY$. It is equal to the angle between the reaction plane P$_r$ and the decay plane
P$_d$.} \label{fig1}
\end{center}
\end{figure}

The general formalism for photoproduction of two pseudoscalar mesons on nucleons has been
developed in Refs.\,\cite{RobertsOed} and \cite{FiAr11}, where the formulas for different
polarization observables are presented. In the expressions below, we denote the final
state particles as 1, 2, and 3  and consider the particle selection $1+(2\,3)$, with
reference to the coordinate system presented in Fig.\,\ref{fig1}. The results are
presented for three independent particle sets, corresponding to the numbering
$1+(2\,3)=\eta+(\pi\,p)$, $\pi+(\eta\,p)$, and $p+(\pi\,\eta)$. In each case the $z$ axis
is directed along the photon momentum. The $x$ and $y$ axes are chosen such that the
momentum of particle 1 has a positive $x$ projection and is orthogonal to the $y$ axes.
As independent kinematical variables we choose angles $\Omega_1=(\Theta_1,\Phi_1=0)$ of
particle 1 in the overall center-of-mass (cm) system, together with angles
$\Omega_{23}^*=(\theta_{23}^*,\phi_{23}^*)$ of particle 2 in the cm system of the
pair $(2\,3)$ and their corresponding invariant mass $M_{23}$.
%============================= Fig. 1 ===============================>
%=====================================================================>

If the target nucleon is transversally polarized and the incident photon beam is
circularly polarized, the cross section can be written in the form (see Eq.\,(57) of
Ref.\,\cite{FiAr11})
\begin{eqnarray}\label{eq1}
&&\frac{d\sigma}{d\Omega_1 dM_{23}d\Omega^*_{23}}= \frac{d\sigma_0}{d\Omega_1
dM_{23}d\Omega^*_{23}}\Big\{1+h P_\odot
I^\odot\nonumber\\
&&\phantom{xxxx}+\frac{1}{\sqrt{2}}P_T\big[P_x\cos\phi-P_y\sin\phi \\
&&\phantom{xxxx}+h P_\odot(P_x^\odot\cos\phi-
P_y^\odot\sin\phi)\big]\Big\}\,,\nonumber
\end{eqnarray}
where $P_{\odot}$ and $P_T$  denote the degree of circular beam and transverse target polarization,
$h= \pm 1$ is the beam helicity, and $\phi$ is the azimuthal angle of the target polarization vector in
a coordinate frame fixed to the reaction plane.
The unpolarized differential cross section is denoted as $\sigma_0$.
The circular photon asymmetry $I^\odot$ has already been
discussed in detail in Ref.\,\cite{Aphi}.
For the asymmetries we have used the notation of Ref.\,\cite{RobertsOed}.
As is evident from Eq.\,(\ref{eq1}), for the totally exclusive fivefold cross
section there are two independent transverse target asymmetries ($P_x$ and $P_y$) and
two independent beam-target asymmetries ($P_x^\odot$ and $P_y^\odot$).
Table\,\ref{ta1} schematically explains how these asymmetries are separated by
a proper variation of the photon and proton polarization parameters.
The observables $P_y$ and $P_x^\odot$ are equivalent to the $T$ and $F$ asymmetries in single
pseudoscalar meson photoproduction.
%========================== table 1 =================================>
\begin{table}
\renewcommand{\arraystretch}{1.0}
\caption{Polarization observables measured in the present work. Notations from
Ref.\,\cite{RobertsOed} are used.} \label{ta1}
\begin{center}
\begin{tabular*}{8cm}%{\textwidth}
{@{\hspace{0.6cm}}c@{\hspace{0.6cm}}|@{\hspace{1.6cm}}c@{\hspace{1.6cm}}c}
\hline%\noalign{\smallskip}
Beam & Target \\
\hline
     & x  &  y \\
\hline%\noalign{\smallskip}
$-$  & $P_x$ & $P_y$  \\
$c$  & $P_x^\odot$ & $P_y^\odot$\\
\hline
\end{tabular*}
\end{center}
\end{table}
%============================ end table =============================>

%%%%%%%%%%%%%%%%%%%%%%%%%%%%%%%%%%%%%%%%%%%%%%%%%%%%%%%%%%%%%%%%%%%%%%%%%%%%%%%%%%%%%%%%%%%%%
\section{Model}\label{sec:model}

For the interpretation of our results we adopted an isobar model approach as used, 
for example, for double pion photoproduction in Refs.\,\cite{Oset,Laget,Ochi,Mokeev,FA}. 
The main ingredients are described in detail in Refs.\,\cite{FOT,FKLO}. 
Here we limit ourselves to
a brief overview needed for the discussion below. 

The reaction amplitude $T$ contains background and resonance terms
\begin{equation}\label{TBR}
T=T^B+\sum_{R(J^\pi;I)}T^R\,,
\end{equation}
where each resonance state $R(J^\pi;I)$ is determined by spin-parity $J^\pi$ and isospin
$I$.

%%%%%%%%%%%%%%%%%%%%%%%%%%%%% Fig. 2 %%%%%%%%%%%%%%%%%%%%%%%%%%%%%%%%%%%
\begin{figure}
\begin{center}
\resizebox{0.47\textwidth}{!}{%
\includegraphics{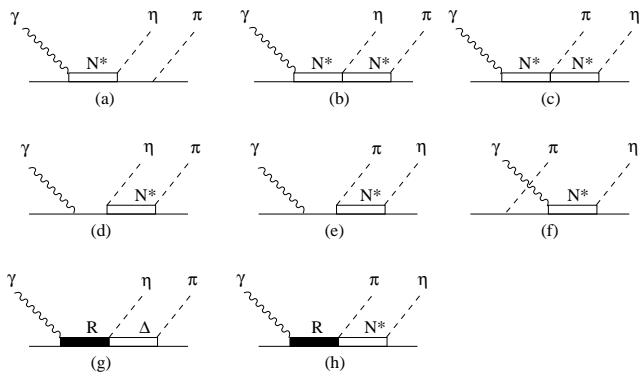}}
\caption{Diagrams representing the amplitude for the $\gamma N\to\pi\eta N$. The
notations $\Delta$ and $N^*$ are used for the resonances $\Delta(1232)$ and
$S_{11}(1535)$.} 
\label{fig_Diagr}
\end{center}
\end{figure}
%%%%%%%%%%%%%%%%%%%%%%%%%%%%%%%%%%%%%%%%%%%%%%%%%%%%%%%%%%%%%%%%%%%%%%%%%

The resonance sector [diagrams (g) and (h) in Fig.\,\ref{fig_Diagr}] includes only the
states with isospin $I = 3/2$. As already noted, analysis of the existing data for
$\gamma p\to\pi^0\eta p$ is in general agreement with the assumption that in the energy
region $E_\gamma < 1.4$ GeV the reaction is dominated by the $D_{33}$ partial wave. In the
present model, the latter is populated by the $\Delta(1700)3/2^-$ and
$\Delta(1940)3/2^-$ states. The resonance $\Delta(1940)3/2^-$ was introduced into the
reaction $\gamma p\to\pi^0\eta p$ in Ref.\,\cite{HornEPJ}. 
In Ref.\,\cite{FKLO} it was needed in order to maintain the importance of the
$D_{33}$ wave at energies above 1.3 GeV, which otherwise would rapidly decrease with
increasing energy. Other $I = 3/2$ resonances entering the amplitude are
$\Delta(1750)P_{31}$, $\Delta(1920)P_{33}$, $\Delta(1600)P_{33}$, and
$\Delta(1905)F_{35}$. 

According to the isobar model concept, each resonance state $R(J^\pi;T)$ generates the
final $\pi\eta N$ state via intermediate transitions to $\eta\Delta(1232)$ and $\pi
S_{11}(1535)$ configurations. In this respect the resonance terms $T^R$ in (\ref{TBR})
are given by a coherent sum of two amplitudes,
\begin{equation}\label{T12}
T^R=T^{(\eta\Delta)}+T^{(\pi N^*)}\,,
\end{equation}
where the isobars $\Delta(1232)$ and $S_{11}(1535)$ are denoted as $\Delta$ and $N^*$,
respectively. Each term in (\ref{T12}) has the form
\begin{eqnarray}\label{Talpha}
&&T^{(\alpha)}=A_\lambda\, G_R(W)\, f^{(\alpha)}(W,\vec{q}_\pi,\vec{q}_\eta,\vec{p}_N)\,,\nonumber\\
&&\alpha=\eta\Delta\,,\ \pi N^*\,.
\end{eqnarray}
with $W$ being the total center-of-mass energy. The quantities $A_\lambda$ are helicity
functions determining the transition $\gamma N\to R$. The propagators $G_R$ were
calculated in the nonrelativistic form
\begin{equation}
G_R(W)=\frac{1}{W-M_R+\frac{i}{2}\Gamma(W)}\,.
\end{equation}
The total energy dependent width $\Gamma$ is a sum of the partial decay widths in $\pi
N$, $\eta\Delta$ and $\pi N^*$ channels:
\begin{equation}
\Gamma=\Gamma_{\pi N}+\Gamma_{\pi\eta N}^{(\eta\Delta)}+\Gamma_{\pi\eta N}^{(\pi N^*)}\,.
\end{equation}
The latter two, $\Gamma_{\pi\eta N}^{(\eta\Delta)}$ and $\Gamma_{\pi\eta N}^{(\pi N^*)}$,
were calculated with explicit inclusion of the finite widths of the $\Delta$ and $N^*$
isobars. Finally, the functions $f^{(\alpha)}$ in Eq.\,(\ref{Talpha}) depending on the
three-momenta of the final particles describe decays of the resonances into the final $\pi\eta
N$ state. As adjustable parameters the
Breit-Wigner masses $M_R$, as well as the products $\sqrt{\Gamma^{(\alpha)}_{\pi\eta
N}}A_\lambda$ ($\alpha=\eta\Delta,\,\pi N^*$), were used in \cite{FKLO}. The total widths of resonances
were not varied. The closeness of the resonances to the
$\pi\eta$ production threshold, especially of $\Delta(1700)D_{33}$, results in rather
weak sensitivity of the cross section to their widths. Therefore the values of the total
widths at the resonance position $\Gamma(M_R)$ were taken directly from the Particle Data
Group (PDG) compilation \cite{PDG} or from the references cited there. For the same
reason, the masses of the resonances, rated by four or three stars, were varied around
their PDG values.

The slowly varying background $T^B$ is presented in Fig.\,\ref{fig_Diagr} by the diagrams
(a) to (f). Only those diagrams were taken into account whose contribution is
appreciable. We have omitted, for example, the terms with $\Delta$ and $N^*$ isobars in
the $u$ channel. The diagrams (b) and (c) contain the unknown coupling constants in the
$\pi N^*N^*$ and $\eta N^*N^*$ vertices. Since the corresponding terms have rather small
impact on the calculation results, these constants were not treated as adjustable
parameters and just for simplicity were fixed according to the prescription
\begin{equation}
f_{\pi N^* N^*}= f_{\pi N N}\,,\quad f_{\eta N^* N^*}= f_{\eta N N}\,.
\end{equation}
As the direct calculation in \cite{Doring,FOT,FKLO} shows, the background terms do not
contribute significantly to the cross section.
%Several comments about the model are in order. Firstly, t
The $\pi\eta$ system is assumed
not to resonate in the energy region considered. The validity of this assumption is
confirmed by the results of Ref.\,\cite{HornEPJ} where the contribution of the resonance
$a_0(980)$ at energies $E_\gamma < 1.4$ GeV is shown to be less than 1$\%$. Furthermore, the
model does not contain relative phases in the electromagnetic couplings which are
sometimes used in the multipole analyses. Although these phases allow one to make the
phenomenological formulation more flexible, their inclusion leads to unnecessary increase
of the number of adjustable parameters.

%%%%%%%%%%%%%%%%%%%%%%%%%%%%%%%%%%%%%%%%%%%%%%%%%%%%%%%%%%%%%%%%%%%%%%%%%%%%%%%%%%%%%%
\section{Experimental setup}

The experiment was performed at the MAMI C accelerator in Mainz\,\cite{MAMIC} using the
Glasgow-Mainz tagged photon facility\,\cite{TAGGER}.
Bremsstrahlung photons were produced by scattering a longitudinally polarized 
electron beam with an energy of 1557 MeV and a polarization degree of 80\% on a 
10-$\mu$m-thick copper radiator.
The photons are energy tagged by momentum analysis of the scattered electrons 
in the dipole magnet spectrometer. 
The resulting energy-tagged photon beam 
covered an energy range from 450 to 1450 MeV with an average resolution of 4 MeV.
The polarization degree of the electron beam was measured
periodically using Mott scattering at the laser-driven source.
The beam helicity was switched randomly, with a frequency of 1 Hz, during the experiment
and the orientation of the polarization vector at the radiator position was checked 
using Moeller scattering. In the Bremsstrahlung process, the longitudinal polarization 
of the electrons is transferred to the circular polarization of the emitted photons \cite{Olsen}.
The degree of circular photon polarization $P_{\odot}$ depends on the photon energy 
and varied from 67\% at 1050 MeV to 79\% at 1450 MeV.

\begin{figure}[t]
\begin{center}
\resizebox{0.38\textwidth}{!}{%
\includegraphics{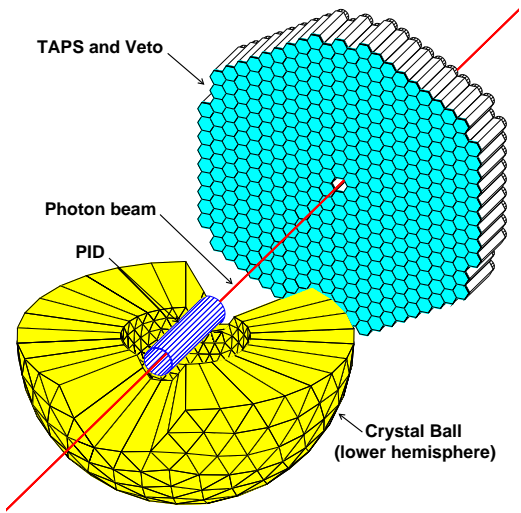}}
\caption{(Color online) Experimental setup with the upper hemisphere of the Crystal Ball 
omitted to show the central region.}
\label{fig2_2}
\end{center}
\end{figure}

The reaction $\gamma p\to \pi^0\eta p$ was measured using the Crystal Ball
(CB) central spectrometer \,\cite{CB} with TAPS \,\cite{TAPS} as a forward angle
spectrometer. The full detector setup 
is shown schematically in Fig.\,\ref{fig2_2}.
%============================= Fig. 3 ===============================>
%=====================================================================>
The spherical CB detector consisted of 672 optically insulated
NaI(Tl) crystals with a thickness of
15.7 radiation lengths pointing towards the center of the sphere. The crystals were 
arranged in two hemispheres covering 93\% of the full solid angle. 
Electromagnetic showers were reconstructed with an energy resolution of  
$\Delta E/ E = 1.7\%$ at 1 GeV. Shower directions were
measured with a resolution of $\sigma_{\theta} \approx 2 - 3^\circ$ in the polar and
$\sigma_{\phi} \approx 2^\circ/\sin \theta$ in the azimuthal angle.
For charged-particle identification via differential energy loss, a barrel of 24 thin 
scintillation detectors surrounding the target was used \cite{PID}.
The forward angular range $\theta = 1 - 20^\circ$ was covered by the TAPS calorimeter \,\cite{TAPS}, 
arranged as a planar configuration of 384 hexagonally shaped BaF$_2$ detectors. 
Each detector had an inner diameter of 5.9 cm and was 25 cm long, which corresponds 
to 12 radiation lengths. The resolutions for the reconstruction of electromagnetic  showers,
were $\sigma/E_{\gamma} = 0.0079/(E_{\gamma}/GeV)^{0.5} + 0.018$ for the energy 
and $\sim 1^\circ$ for the direction. A 5-mm thick plastic scintillator in front of each module 
allows the separation of neutral and charged particles.
Photons (or electrons) and hadrons can be separated by a pulse-shape analysis
based on the properties of $BaF_2$. The crystals  have the fast and long components of
the scintillation, the intensity of which depends on the incident particle nature.
Analysis of these components gives us an additional method of particle identification.
The best way to identify the charged particle species in TAPS is a time-of-flight versus
cluster energy analysis. 
The solid angle of the combined Crystal Ball and TAPS detection system is nearly
$97\%$ of $4\pi$ sr.

The transversely polarized target protons were provided by a
frozen-spin butanol ($\mathrm{C_4H_9OH}$) target \cite{Thomas}.  
A four-layer saddle coil provided a 0.45~T holding field perpendicular to the beam
axis at a current of 35 A.
A $\mathrm{^3He/^4He}$ dilution refrigerator keeps the target material at a temperature 
of 25 mK which corresponds to relaxations times of 1500~h.
The 2-cm-long and 2-cm-diameter cylindrical target cell was filled with 2-mm-diameter butanol
spheres with a packing fraction (filling factor) of $\sim 60\%$.
The target polarization was measured using the NMR techniques at the beginning and the end
of each data taking period. The polarization was then calculated for each individual
data file from the known exponential relaxation of the polarization.
In order to reduce the systematic uncertainties, the direction of the target polarization vector
was regularly reversed during the experiment.
The average degree of polarization during the beam periods May-June 2010 and April 2011
was $70\%$.

%%%%%%%%%%%%%%%%%%%%%%%%%%%%%%%%%%%%%%%%%%%%%%%%%%%%%%%%%%%%%%%%%%%%%%%%%%%%%%%%%%%%%%
\section{Data analysis}\label{Data}

\begin{figure}
\begin{center}
\resizebox{0.47\textwidth}{!}{%
\includegraphics{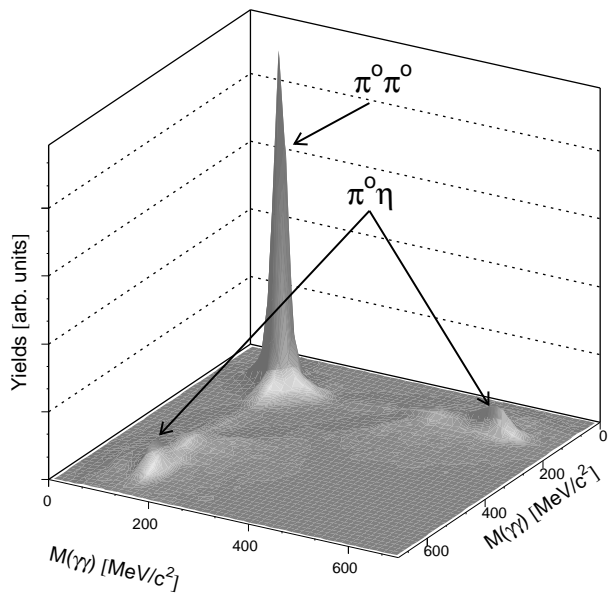}}
\caption{Event selection for final states with four photons: $M_{\gamma\gamma}$ {\it vs}
$M_{\gamma\gamma}$ for all possible independent combinations of $\gamma\gamma$ pairs (three
entries for each event).}
\label{fig2}
\end{center}
\end{figure}

The reconstruction of the $\gamma p\to \pi^0 \eta p$ reaction is based on 
the two photon decays of the $\pi^0$ and the $\eta$ meson as described in 
detail in Ref.\,\cite{PiEtaEPJA}.
As a first step, events with four neutral and one or zero charged particles in the Crystal Ball
and TAPS detectors were selected. 
The distribution of invariant masses, calculated from all possible
combinations of the four neutral hits is shown in Fig.\,\ref{fig2}. As there are three independent combinations of
possible pairs, this histogram has three entries per event. The distribution shows 
already large peak corresponding to the $\pi^0\pi^0$ channel and two smaller ones from 
the $\pi^0\eta$ final state. 
In the next step, a $\chi^2$ for both possible final
states, $\pi^0 \pi^0$ and $\pi^0\eta$, was calculated for each possible
permutation of the four  neutral hits: 
%========================================>
\begin{eqnarray}
\chi^2_{2\pi}&=&
\left(\frac{M_{\gamma_i\gamma_j}-m_{\pi^0}}{\sigma_{\pi^0}}\right)^2 +
\left(\frac{M_{\gamma_k\gamma_l}-m_{\pi^0}(m_{\eta})}{\sigma_{\pi^0}(\sigma_{\eta})}\right)^2. \label{chi1} 
\end{eqnarray}
%========================================>
Here $m_{\pi^0}$ and $m_\eta$ are $\pi^0$ and $\eta$ masses and $\sigma_{\pi^0} = 10$ MeV
and $\sigma_\eta = 25$ MeV are the corresponding invariant mass resolutions of the
detector system. Each event was then assigned to either $\pi^0\pi^0$ or $\pi^0\eta$
production depending on the minimum of the $\chi^2$ values.
Further selection is based on selections in the invariant $M(\gamma \gamma)$ mass
distributions and the $MM(\gamma, \pi^0 \eta)$ missing mass calculated with the assumption
of a $\gamma p$ initial state and the reconstructed $\pi^0 \eta$ pair.
%============================= Fig. 4 ===============================>
%====================================================================>
%============================= Fig. 5 ===============================>
\begin{figure}
\begin{center}
\resizebox{0.48\textwidth}{!}{%
\includegraphics{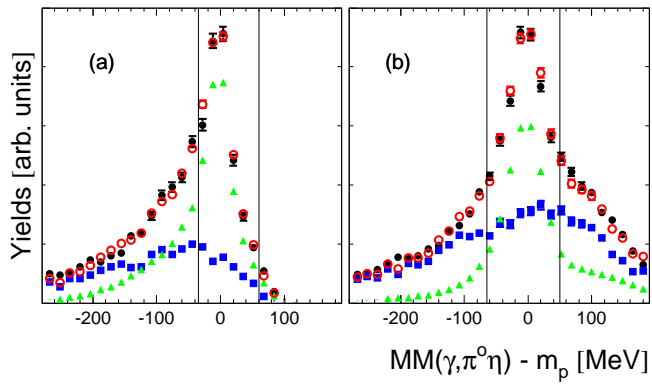}}
\caption{(Color online) Missing mass distributions corresponding to photon 
beam energies of (a) 1100 MeV  and (b) 1400 MeV.
The full black circles are obtained with butanol data. 
The green triangles and blue squares are hydrogen
and carbon data scaled to fit the butanol data. The fit result, which is a sum of the hydrogen
and carbon partial contributions, is shown by the open red circles.}
\label{fig3}
\end{center}
\end{figure}
%====================================================================>

In principle, the polarization observables  in Eq. (1)
can be determined in each photon energy, angular or invariant mass $M_{23}$ 
bin as count rate asymmetries from the number $N^{\pm}$ of reconstructed
$\vec \gamma \vec p \to \pi^0\eta p$ events with different
orientations of target spin and beam helicity:
\begin{equation}\label{NPx}
P_x =\frac{1}{P_T |\cos\phi|}\,\frac{N^{\pi = +1}-N^{\pi = -1}}{N^{\pi = +1}+N^{\pi = -1}}\,,
\end{equation}
\begin{equation}\label{NPy}
P_y =\frac{1}{P_T |\sin\phi|}\,\frac{N^{\pi = +1}-N^{\pi = -1}}{N^{\pi = +1}+N^{\pi = -1}}\,,
\end{equation}
\begin{equation}\label{NPxc}
P_x^\odot =\frac{1}{P_T |\cos\phi|}\,\frac{1}{P_\odot}\,
\frac{N^{\sigma = +1}-N^{\sigma = -1}}{N^{\sigma = +1}+N^{\sigma = -1}}\,,
\end{equation}
\begin{equation}\label{NPyc}
P_y^\odot =\frac{1}{P_T |\sin\phi|}\,\frac{1}{P_\odot}\,
\frac{N^{\sigma = +1}-N^{\sigma = -1}}{N^{\sigma = +1}+N^{\sigma = -1}}\,,
\end{equation}
where $\pi = \vec p_T \cdot \hat y/|\vec p_T \cdot \hat y| = \pm 1$ denotes the orientation of the target
polarization vector
$\vec p_T$ relative to the normal of the production plane and
$\sigma = h \; \vec p_T \cdot \hat x/|\vec p_T \cdot \hat x| = \pm 1$ is given by the product of the beam
helicity $h$
and the orientation of $\vec p_T$ relative to the $x$ axis.
In these asymmetries, systematic uncertainties
related to the total photon flux normalization and the target filling factor
cancel. 
However, using a butanol target has one essential disadvantage due to additional
background from reactions on $\mathrm{^{12}C}$ and $\mathrm{^{16}O}$ nuclei.
In the numerators of Eqs.~(\ref{NPx})-(\ref{NPyc}), this background cancels because the nucleons bound in
$\mathrm{^{12}C}$ or $\mathrm{^{16}O}$ are
unpolarized. However, in order to determine the denominator, this contribution has
to be taken into account.
The detection of the outgoing protons and applying kinematic constraints already suppress this
background significantly.
In order to subtract the remaining background we analyzed $\pi^0\eta$ photoproduction on
pure carbon and liquid hydrogen targets.
The corresponding $MM(\gamma, \pi^0 \eta)$ distributions were scaled and added
in order to fit the corresponding distribution obtained with the butanol target.
Since the magnitude and the shape of the background depend on the initial beam
energy and on the momenta of the final particles, this procedure was
performed for each individual kinematical bin.
This subtraction method is illustrated on Fig.\,\ref{fig3} for
two photon energies, which are typical for the presented data analysis. Missing mass
spectra for the
reaction $\gamma p\to \pi^0\eta p$ with the butanol target are shown
by the full black circles. Spectra measured with the hydrogen and carbon
targets are represented by the green triangles and the blue squares
respectively. Their absolute values were scaled to fit the butanol data. The red
open circles represent the sum of the fitted hydrogen and carbon contributions.
For further analysis only events around the proton peak, within the
vertical lines, are used. The intervals were selected to optimize the 
signal-to-background ratio. 
Even in the distribution from the
pure hydrogen (green triangles in Fig.~\ref{fig3} still some background remains. 
This is mainly due to misidentified $\gamma p\to \pi^0\pi^0 p$ reactions, split or overlapping
photon clusters, and combinatoric mixing. 
This background is subtracted by fitting the missing
mass distributions with the sum of a Gaussian and a third-order polynomial 
function. After subtracting the polynomial background, the distribution is
found to be in excellent agreement with results of a Monte Carlo simulation 
of the $\gamma p \to \pi^0 \eta p $ reaction (see Fig. 4 in Ref.~\cite{PiEtaEPJA}).
The average detection efficiency for the $\gamma p \to \pi\eta p$ reaction
after all analysis steps is about 50\% for beam energies from 1.05 to 1.45 GeV.

\begin{figure}
\begin{center}
\resizebox{0.47\textwidth}{!}{%
\includegraphics{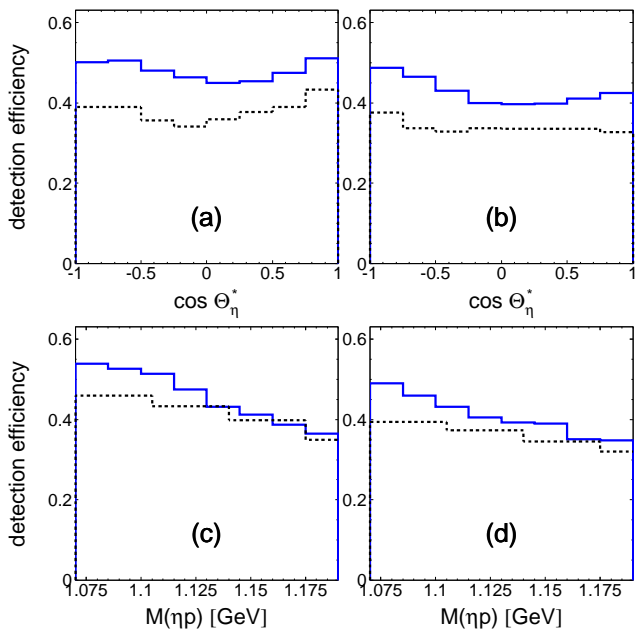}}
\caption{(Color online) Detection efficiency at fixed $\Theta_{\pi}$ for two photon energy bins: $1100
\pm 50$ MeV (solid blue line) and $1400 \pm 50$ MeV (black dashed line). $\Theta^*_\eta$ is
the polar angle in the cm system of the $(\eta p)$ pair. Panels (a) and (c) show the
efficiency at $\cos \Theta_{\pi} = -0.9 \pm 0.1$, and (b) and (d) at $\cos \Theta_{\pi} =
0.9 \pm 0.1$.}
\label{fig_Eff}
\end{center}
\end{figure}

The asymmetries in each photon energy, angular or invariant mass $M_{23}$ bin
were obtained as count rate asymmetries by integrating the reaction yields in 
Eqs.~(9)-(12) over four remaining variables.
This procedure is exactly valid if the acceptance does not depend 
on any of these variables. However, a small and smooth variation of the 
acceptance can be taken into account in the systematic uncertainties. 
Figure \ref{fig_Eff} shows as an example the efficiency as function of 
those kinematic variables where the strongest variation of the acceptance 
is observed. The typical variation is less then 3\%. 
The influence of the acceptance variations and background subtraction on the 
asymmetries was estimated by varying the applied selection criteria. The observed 
changes in the asymmetries by 3\%-6\%  were smaller than the statistical uncertainty. 
Other systematic uncertainties of the present measurement are related to the
determination of the proton polarization (4\%) and the beam polarization (2\%). 
By adding all contributions in quadrature, a total systematic uncertainty of less
than 8\% was obtained.

%============================= Fig. 6 ===============================>
%====================================================================>

%%%%%%%%%%%%%%%%%%%%%%%%%%%%%%%%%%%%%%%%%%%%%%%%%%%%%%%%%%%%%%%%%%%%%%%%%%%%%%%%%%%%%%%%%%%%%
\section{Discussion of the results}\label{sec:discussion}

\begin{figure*}
\begin{center}
\resizebox{0.9\textwidth}{!}{%
\includegraphics{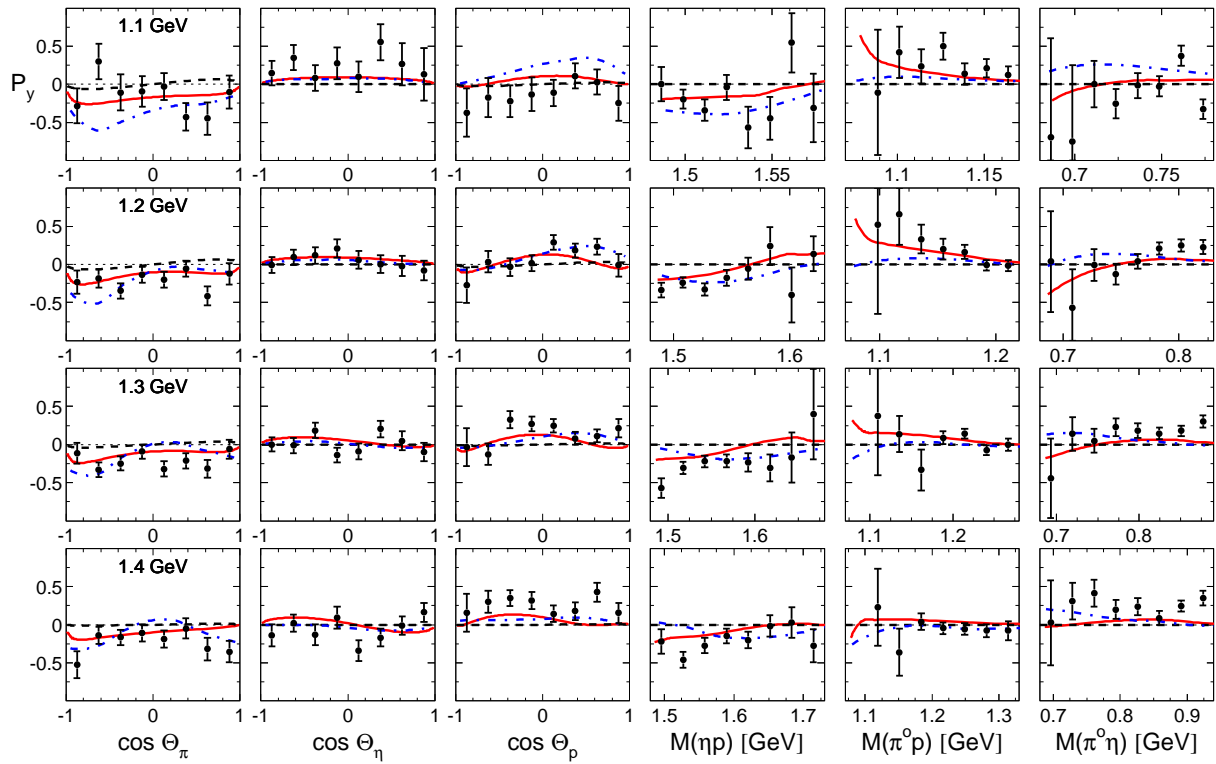}}
\caption{(Color online) Angular and invariant mass distributions for the target
asymmetry $P_y$ of the $\gamma p\to\pi^0\eta p$ reaction for incident photon
energies from 1050 to 1450 MeV.
Our experimental results with statistical uncertainties are shown by filled circles.
The solid curves show the prediction of the isobar model \protect\cite{FKLO}. The dashed
curves include only the $3/2^-$ partial wave.
Predictions of the Bonn-Gatchina model \protect\cite{Anis12} are shown by
dashed-dotted curves.
The energy labels on the left panels indicate the central energy of
the four 100-MeV-wide photon energy bins.}
\label{fig4}
\end{center}
\end{figure*}

\begin{figure*}
\begin{center}
\resizebox{0.9\textwidth}{!}{%
\includegraphics{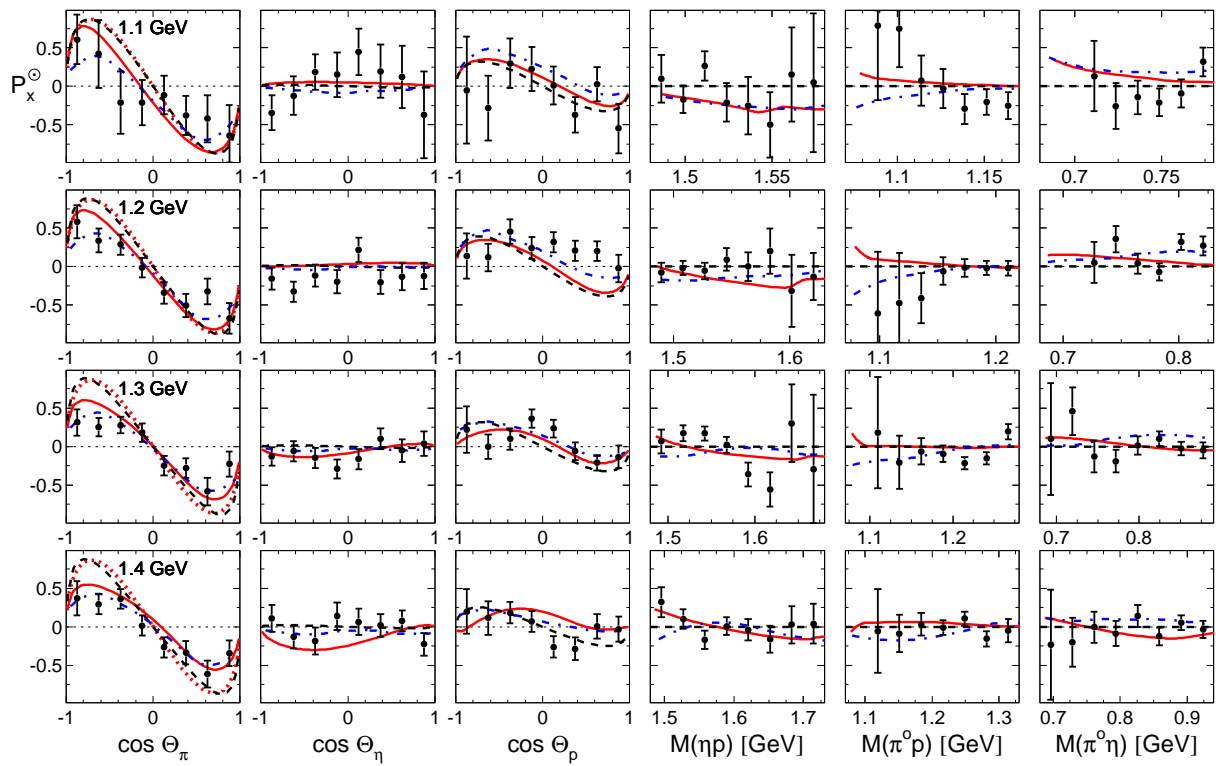}}
\caption{(Color online) Same as Fig.\,\protect\ref{fig4} for the beam-target asymmetry
$P_x^\odot$. The red dotted line in the first column was obtained using Eq. (7).
}
\label{fig5}
\end{center}
\end{figure*}

\begin{figure}
\begin{center}
\resizebox{0.48\textwidth}{!}{%
\includegraphics{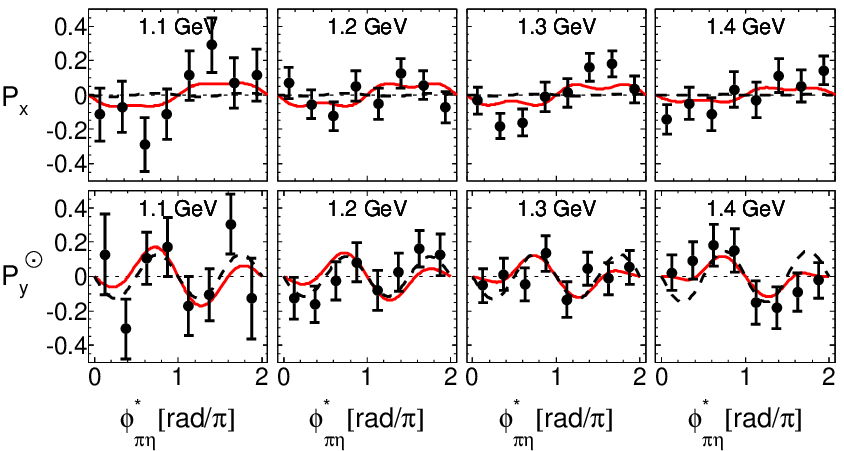}}
\caption{(Color online) The target asymmetry $P_x$ and
beam-target asymmetry $P_y^\odot$ as functions of $\phi_{\pi\eta}^*$.
All notation is the same as in Fig.\,\protect\ref{fig4}.}
\label{fig6}
\end{center}
\end{figure}

Figures \,\ref{fig4}, \ref{fig5}, and \ref{fig6} show the measured asymmetries as function of the
various scattering angles and invariant mass combinations.
%============================= Fig. 7 ===============================>
%====================================================================>
%============================= Fig. 8 ===============================>
%====================================================================>
%============================= Fig. 9 ===============================>
%=====================================================================>
Data from the $\pi+(\eta p)$ analysis are shown in columns 1 and 4. The $\eta+(\pi p)$ and $p+(\pi \eta)$
data are shown in columns $2 and 5$ and in columns $3 and 6$, respectively.
In this section we present
a qualitative description of the reaction properties revealed by our data.
A detailed partial wave analysis of the new data has started and will be published elsewhere.

As one can see in Figs.\,\ref{fig4} and \ref{fig5}, both $P_y$ and $P_x^{\odot}$
have rather small values, except for $P_x^{\odot}(\cos\Theta_\pi)$ which
strongly varies with pion angle (see first column in Fig.\,\ref{fig5}). This dependence can
be described to a
good approximation by an odd function with a maximum amplitude of about 0.8. This
peculiar behavior is a direct signature of $s$-wave production of the $\eta\Delta$
configuration in the $J^\pi=3/2^-$ state. As noted in Sec.\,\ref{Intro} this
partial wave appears to dominate the reaction amplitude in a wide energy region from
threshold to $E_\gamma=1.5$ GeV \cite{Horn1940,FKLO}. Assuming that the reaction proceeds
exclusively via the formation of an intermediate $\eta\Delta$ pair in the $3/2^-$ partial wave,
one obtains the following simple form for $P_x^\odot$ in the region of $\Delta(1700)D_{33}$:
\begin{equation}\label{Func}
P_x^\odot\approx -\frac{2}{\sqrt{3}} \frac{A_{1/2}A_{3/2}\sin
2\Theta_\pi}{A_{1/2}^2(1/3+\cos^2\Theta_\pi)+A_{3/2}^2\sin^2\Theta_\pi}\,,
\end{equation}
where $A_\lambda$ ($\lambda=1/2,\,3/2$) are  the helicity amplitudes for the transition
$\gamma N\to\Delta(1700)D_{33}$. If we further assume that $A_{1/2}\approx A_{3/2}$
\cite{PDG} Eq.~(13) reduces to the simpler approximation
\begin{equation}\label{Func1}
P_x^\odot\approx -\frac{\sqrt{3}}{2}\sin 2\Theta_\pi\,.
\end{equation}
The function (\ref{Func1}) reaches its maximum value $P_x^\odot=\sqrt{3}/2$ at
$\cos\Theta_\pi=-1/\sqrt2$, in general agreement with our data (see red dotted lines
in the first column in Fig. \ref{fig5}).

It is also worth noting that, in the same $s$-wave hypothesis of $\eta\Delta$ production,
the behavior of $P_x^\odot(\cos\Theta_\pi)$ should be similar (up to the possible change
of sign) to that of the observable $F$ for single $\pi^0$ photoproduction in the $\Delta$
resonance region. Indeed, assuming that the $\Delta(1232)$ excitation is a pure magnetic
dipole transition the distribution $F(\cos\Theta)$ has the form
\begin{equation}
F\approx \frac{3\sin2\Theta}{5-3\cos^2\Theta}\,,
\end{equation}
so that, as in the case of $P_x^\odot$ [see Eq. (13)], the angular dependence of
$F$ is mainly governed by the factor $\sin2\Theta$.

Concerning the role of positive parity states, two important facts can be observed.
First, the dependence of $P_y$ and $P_x^\odot$ on the invariant mass $M_{23}$
is determined exclusively by an interference of partial wave amplitudes with
opposite parities. The simple model with only the $\Delta 3/2^-$ amplitude
therefore gives the trivial result
\begin{equation}\label{eq4}
d{\cal O}/dM_{23}=0\,,\quad {\cal O}=\{P_y,P_x^\odot\}\,.
\end{equation}
Second, if only the dominant $\Delta 3/2^-$ wave is included, both $P_y$ and
$P_x^\odot$ are odd functions of $\Theta_1-\pi/2$
\begin{equation}\label{eq8}
{\cal O}(-\cos\Theta_1)=-{\cal O}(\cos\Theta_1)\,,\quad {\cal O}=\{P_x,P_y^\odot\}\,,
\end{equation}
and reach zero at $\Theta_1=\pi/2$. In this respect, the nonzero values of $P_y$ and
$P_x^\odot$ at $\Theta_1=\pi/2$ as well as of their distribution over the invariant
mass $M_{23}$ may be viewed as a signature of the presence of partial waves with
positive parity. However, as evident from Figs.\,\ref{fig4} and \ref{fig5}
the deviation of the measured values from the simple rules (\ref{eq4}) and (\ref{eq8}) is small,
indicating that the role of states besides $\Delta 3/2^-$ is not large.
This is in full agreement with our previous results for the unpolarized angular distribution
\cite{W_LM}  as well as for the helicity beam asymmetry \cite{Aphi}.

The solid lines in Figs.\,\ref{fig4}-\ref{fig6} show the prediction of the isobar model
described in Sec.\,\ref{sec:model}.
Here we use the parameter set (I) (see Table I in Ref.\,\cite{FKLO}) which was preferred
since it gives the best description of the measured linear beam asymmetry $\Sigma$
\cite{Ajaka,Gutz1}. This solution also reproduces the invariant mass distributions
measured in \cite{Ajaka} (see Fig.\,9 in Ref.\,\cite{FKLO}) and describes reasonably well
the data for the beam asymmetries $I^{\odot}$, $I^c$ and $I^s$ presented in
Refs.\,\cite{Aphi,Gutz2}.

The dash-dotted lines in Figs.\,\ref{fig4} and \ref{fig5} show predictions of the
Bonn-Gatchina multichannel fit \cite{Anis12} (solution BG2011-02). In contrast to
\cite{FKLO}, where only the data for $\gamma p\to\pi^0\eta p$ were fitted, within the
Bonn-Gatchina approach, the positions of resonances, their partial decay widths, and
relative strengths were fitted simultaneously to the data sets in different channels,
including single and double meson production as well as strangeness production. The
application to the reaction $\gamma p\to\pi^0\eta p$ is described in detail in
Ref.\,\cite{HornEPJ}. In this analysis some contributions from $N^*$ resonances, which do
not enter the amplitude in \cite{FKLO}, in particular the $N(1880)P_{11}$, are also
included.

Both models describe the new data equally well. The present statistics do not allow any
discrimination between the different model predictions in kinematic regions where they
show small differences, e.g., at low values of $M(\pi^0p)$.

The other two observables $P_x$ and $P_y^\odot$ contribute exclusively to the
distribution of the cross section over the azimuthal angle $\phi_{23}^*$ and vanish in
the distribution over $\Theta_1$ and $M_{23}$. In Fig.\,\ref{fig6} we show
data for the particle selection $1+ (2 3) = p + (\pi^0 \eta)$.
As in Ref.\,\cite{Aphi} the denominators of the asymmetries are averaged
over the whole $\phi_{\pi\eta}^*$ region.
Parity conservation requires
\begin{equation}\label{eq10}
{\cal O}(\phi^*_{23})=-{\cal O}(2\pi-\phi^*_{23})\,,\quad {\cal O}=\{P_x,P_y^\odot\}\,.
\end{equation}
Using angular momentum algebra it can be shown that if only the states with $J^P=3/2^-$
enter the amplitude (in our case $\Delta3/2^-$ resonances) the product ${\cal
O}d\sigma_0/d\phi^*_{23}$, for both ${\cal O}=P_x$ and ${\cal O}=P_y^\odot$, is
proportional to $\sin 2\phi_{\pi\eta}^*$ and does not contain higher order harmonics.
The presence of states with positive parity leads to a more complicated shape for these
observables, as observed in our data.

%%%%%%%%%%%%%%%%%%%%%%%%%%%%%%%%%%%%%%%%%%%%%%%%%%%%%%%%%%%%%%%%%%%%%%%%%%%%%%%%%%%%%%
\section{Summary and conclusions}\label{conclusion}

We have presented the first experimental results for the target and the beam-target asymmetries
of the $\gamma p\to\pi^0\eta p$ cross section obtained with circularly polarized photons
and transversely polarized protons.
The measurements were performed using Crystal Ball and TAPS spectrometers.
We presented a qualitative analysis which shows that the new data for all four observables are
in broad agreement with the dominance of the $\Delta 3/2^-$ amplitude, confirming the theoretical
interpretation of previous measurements \cite{Aphi,W_LM} and other
analyses \cite{HornEPJ,FKLO}, in the region below $E_\gamma=1.5$\,GeV. However, the detailed
distributions of the measured observables are sensitive to the contribution of small components
in the reaction amplitude. Specifically an interference between $\Delta 3/2^-$
and the positive parity amplitudes $\Delta 1/2^+$ and $\Delta 3/2^+$ is responsible for the
nontrivial
angular and energy dependence of the asymmetries presented.

%%%%%%%%%%%%%%%%%%%%%%%%%%%%%%%%%%%%%%%%%%%%%%%%%%%%%%%%%%%%%%%%%%%%%%%%%%%%%%%%%%%%%%
\section*{Acknowledgments}

The authors wish to acknowledge the excellent support of the accelerator group of MAMI.
This material is based upon work supported by 
the Deutsche Forschungsgemeinschaft (SFB 443, SFB 1044); 
the European Community Research Activity under the FP7 program (Hadron Physics, Contract No. 227431); 
Schweizerischer Nationalfonds; 
the UK Sciences and Technology Facilities Council (STFC 57071/1, 50727/1); 
the U.S. Department of Energy (Offices of Science and Nuclear Physics, Grants No. DE-FG02-99-ER41110,
No. DE-FG02-88ER40415, and No. DE-FG02-01-ER41194);
the National Science Foundation (Grant No. PHY-1039130, IIA-1358175); 
NSERC (Canada); 
the Dynasty Foundation; 
TPU Grant No. LRU-FTI-123-2014; 
and the MSE Program ``Nauka'' (Project No. 3.825.2014/K).

%%%%%%%%%%%%%%%%%%%%%%%%%%%%%%%%%%%%%%%%%%%%%%%%%%%%%%%%%%%%%%%%%%%%%%%%%%%%%%%%%%%%%%

\end{document}